# Assessing the Sensitivities of Input-Output Methods for Natural Hazard-Induced Power Outage Macroeconomic Impacts

Matthew Sprintson and Edward Oughton

Abstract

It is estimated that over one-fourth of US households experienced a power outage in 2023, costing on average US $150 Bn annually, with 87% of outages caused by natural hazards. Indeed, numerous studies have examined the macroeconomic impact of power network interruptions, employing a wide variety of modeling methods and data parameterization techniques, which warrants further investigation. In this paper, we quantify the macroeconomic effects of three significant natural hazard-induced US power outages: Hurricane Ian (2022), the 2021 Texas Blackouts, and Tropical Storm Isaias (2020). Our analysis evaluates the sensitivity of three commonly used data parameterization techniques (household interruptions, kWh lost, and satellite luminosity), along with three static models (Leontief and Ghosh, critical input, and inoperability Input-Output). We find the mean domestic loss estimates to be US $3.13 Bn, US $4.18 Bn, and US $2.93 Bn, respectively. Additionally, data parameterization techniques can alter estimated losses by up to 23.1% and 50.5%. Consistent with the wide range of outputs, we find that the GDP losses are highly sensitive to model architecture, data parameterization, and analyst assumptions. Results sensitivity is not uniform across models and arises from important *a priori* analyst decisions, demonstrated by data parameterization techniques yielding 11% and 45% differences within a model. We find that the numerical value output is more sensitive than intersectoral linkages and other macroeconomic insights. To our knowledge, we contribute to literature the first systematic comparison of multiple IO models and parameterizations across several natural hazard-induced long-duration power outages, providing guidance and insights for analysts.

## Section 1: Introduction

Power outages ("blackouts") are defined as the unexpected inability of an electrical system to deliver power to a network as a direct consequence of an endogenous or exogenous action associated with the system (IEEE, 1972; Lindstrom & Hoff, 2020). In the short run, power outages can interfere with critical infrastructure systems that rely on electrical power as a core input (transportation, telecommunications, water, etc.) (Aslani et al., 2024; Oughton et al., 2018; Verschuur et al., 2024). In the long run, blackouts can limit consumer access to businesses in physical and online mediums, force businesses to dispose of goods due to lack of maintenance or quality loss, and disrupt supply chains (FEMA, 2024; Koks et al., 2019). Electricity is also a vital factor in the functioning of the digital economy and in private household use, as consumer final demand in the electricity market is predicted to grow to 24% by 2040 (IEA, 2020). The industry provides 12.8% of private value-added but is responsible for 36.8% of supply chain losses due to power outages (Thomas & Fung, 2022). Additionally, power outages have a statistically significant effect on the downstream supply chain for manufacturing and output (Thomas & Fung, 2022).

In 2023, 33.9 million households (.25% of households) experienced at least one power outage, with 70% of those households reporting interruptions of 6 hours or longer (Madamba, 2024). Blackouts are estimated to cost the United States at least US $150 Bn annually and are highly correlated with extreme weather events and specific geographic areas (Do et al., 2023; EIA, 2024a). An analysis found that extreme weather, natural disasters, and other storms cause 86.6% of US power outages. Aging infrastructure and power systems are responsible for 12%, while cyberattacks and other interruptions accounted for the remaining 1.5% (*The Impact of Power Outages*, 2023). Therefore, power outages caused by weather disasters are responsible for a considerable economic impact on American businesses and consumers.

As a result, simulating and measuring the macroeconomic impact of power network interruptions has contributed to insights and recommendations for policy changes (Raza et al., 2022; Sanstad et al., 2020; Stankovski et al., 2023). Analysts use the macroeconomic impact of outages to inform recommendations about risk, resilience, and preparedness (Capitanescu, 2023; Majchrzak et al., 2021; Shrestha et al., 2023). Consequently, macroeconomic modeling of supply chain failures resulting from power outages has been featured in policy discussions and recommendations (Federation of American Scientists, 2024; United States House Committee on Oversight and Accountability, 2024).

With this context in mind, this paper focuses on several wide-area power outages in the United States, including those following Hurricane Ian (2022), the 2021 Texas Blackouts, and Tropical Storm Isaias (2020). The paper evaluates the performance of three static Input-Output (IO) models, including the Leontief and Ghosh models, the Critical Input IO model, and the Inoperability IO model. Then, the paper assesses the sensitivity of each model to data selection methodology by employing three data parameterization techniques, including household interruptions, kWh lost, and satellite luminosity data. IO Methodology has been shown to be a reliable systemic architecture for analyzing the macroeconomic effects of power outages through GDP loss and intersectoral linkages (Macmillan et al., 2023). While static IO modeling has considerable limitations, static IO techniques are frequently employed in analysis and have been shown to provide worthwhile policy recommendations and macroeconomic analysis (Altimiras-Martin, 2024; Awasthi & Sangal, 2025; Ebrahimnejad et al., 2021; Zhu et al., 2025). Finally, the study compares the results of IO analysis with case study data, which has been outlined as an area of future work in IO literature (Macmillan et al., 2023). The research questions include the following:

1. What are the GDP impacts of several recent large-scale power outages in the United States?

2. How do alternative data parameterization methods of natural hazard-induced outages compare with empirical and observed losses?

In Section 2, a literature review examines the macroeconomic impact of power outages and IO modeling procedures for outages. Afterward, a method is presented in Section 3. Modeling results and analysis will be presented in Section 4. We then return to the research questions in Section 5 to discuss possible modeling and policy implications within a broader context. Finally, the conclusions, contributions, and limitations of the study will be presented in Section 6.

# Section 2: Literature Review

In this section, a literature review is conducted to provide insights into the effects of outages on GDP, productivity, and the macroeconomic supply chain. Afterward, a literature review is carried out for IO approaches to analyze the macroeconomic effect of blackouts.

## Section 2.1: The Macroeconomic Effect of Power Outages

Literature has found that power outages have significant macroeconomic effects, threatening final good production, power systems, and the macroeconomic supply chain (Ahmed et al., 2023; Casey et al., 2020; Oughton, Ralph, et al., 2019; T. Huang et al., 2014; Oughton et al., 2017; Oughton, Hapgood, et al., 2019; Kim et al., 2023). One research team found that power outages negatively affect economic growth, with a 1% decrease in the System Average Interruption Duration Index potentially leading to a 2.16% increase in global economic growth. Additionally, the study argued that improving the quality of power infrastructure could improve inequality, narrow wealth disparities, and increase output (Chen et al., 2023; Sprintson & Oughton, 2023). Other research has found that power outages have an adverse effect on supply chains, as some sectors experience up to 36.8% of supply chain losses and lose up to US $133 Bn each year from outages (Ghodeswar et al., 2025; Thomas & Fung, 2022). Indeed, research has shown that outages cause significant direct and indirect economic losses (Oughton et al., 2017), including losses of up to 0.49% of China's national GDP, and severely impact industries with high energy consumption, such as manufacturing (Sun et al., 2024; Q. Wang et al., 2024).

One study employed a simulation to model the economic effects of the 2011 Great East Japan Earthquake. Using survey data, the model incorporated a four-level scale for production disruption in affected regions. The model then assumed that factories affected by power outages shut down production and used the survey to determine the duration of the shutdown. Then, the model accounted for recovery, substitution, and other behavioral effects using probabilistic calculations with respect to location, production capacity, and seismological data. By comparing with a simulation excluding the power outage, the model found that the blackout could have led to a loss of 2.7 trillion yen (Inoue et al., 2023).

Power outages have been found to threaten vital elements of the supply chain, including food access, water availability, infrastructure, and economic stability (Blouin et al., 2024; Pan et al., 2024; Wethal, 2023). Moreover, one study found that the estimated annual cost of US power interruptions is between $35 and $50 Bn with 90% confidence (LaCommare et al., 2018).

One research team studied the impacts of the 2005 Gudrun storm on power service. The researchers used a customer interruption cost equation, which included the monetary value of power interruptions, the value of unexpected outages, and consumers' total energy consumption. The study found that the 2005 blackout cost the Swedish Economy around EUR €3 Bn (Gündüz et al., 2017). Another study used an economic impact indicator incorporating macroeconomic data for industry output, electricity grid, market value, and economic vulnerability. The researchers found that in extreme scenarios like Hurricane Harvey (2017) and Hurricane Irma (2017), the shutdown of the Gowanus Power Plant system could have an economic impact of US $78 Bn and US $68 Bn in Manhattan and Brooklyn, respectively (Garcia Tapia et al., 2019).

To conclude, there is strong evidence of power outages affecting a broad assortment of macroeconomic metrics. Indeed, power disruptions can negatively impact the macroeconomic supply chain and restrict access to an essential input good, thereby decreasing production across many sectors that rely on electricity.

## Section 2.2: Static Input-Output Methodology in Measuring Impacts of Extreme Weather Events

Researchers use static IO modeling methodology to analyze and quantify the macroeconomic effects of natural disasters, such as changes in sectoral output, GDP, and employment (Bella et al., 2023; Lyu et al., 2023; Santos et al., 2023). Static IO modeling methods are most applicable in estimating the indirect effects of disaster-related environmental stimuli from direct effects (R. Huang et al., 2022; Jin et al., 2020). Through IO methods, analysts can estimate a disaster's impact on sectoral final demand to model the final output of each sector in the economy (Akbari et al., 2023; Henriques & Sousa, 2023; Tian et al., 2023).

One study used IO methodology to estimate the macroeconomic impact of the 2021 Texas Severe Weather Storm. The researchers used an inoperability IO model, which estimates the expected GDP loss by augmenting it with a relative sectoral production reduction. Consequently, the paper estimated that the inoperability for the Texan utilities sector was 0.51% and the loss was US $664 Mn. The paper found that oil and gas extraction contributed up to 18% of the total GDP loss of the storm. The utilities industry contributed 11%, and petroleum and coal products contributed 7% (Bhattacharyya & Hastak, 2022). Another study also used IO analysis and found that 1% inoperability in the utilities sector could reduce annual GDP by up to US $11.6 Bn, affecting Utilities (US $1.2 Bn loss), Professional, Scientific, and Technical Services (US $0.8 Bn), and Wholesale Trade (US $0.6 Bn) (Bhattacharyya et al., 2021).

One research team used a risk-based IO framework to estimate the GDP loss of the 2003 Northeast Blackout. The researchers used an inoperability model by normalizing production loss and employing a method similar to Ghosh supply-side methodology to measure downstream losses. The research team curated perturbations by collecting state-level impact data for each day of the outage and calculated a sectoral inoperability of 26%, 20%, and 7% for days 1, 2, and 3, respectively. The research team found that the outage caused a three-day loss of up to US $6.53 Bn, with US $2.12 Bn lost from direct electric power perturbation. The most affected sectors were business services, electricity, and financial institutions (Anderson et al., 2007).

One research team used Ghosh modeling to study the effects of electricity supply disruptions in Iran. The researchers found that a 30% electricity shortage could increase costs by 176% across the Iranian economy. The industries most affected are manufacturing and transportation, with a 42.9% and 24% output loss, respectively (Faridzad et al., 2022). Research has also shown that Ghosh modeling has considerable potential in representing the structure of the economy and deepening structural analysis of macroeconomic stimulus (Altimiras-Martin, 2024; Qu et al., 2018).

One study analyzed several models that estimate the costs of electrical outages, including IO models. The study posited that IO models typically produce an upper bound for outages, with these methods modeling outages lasting several days more accurately. The researchers concluded that simulation-based IO methodology has not been compared to empirical estimates or case studies to evaluate accuracy, reliability, or consistency (Macmillan et al., 2023).

Several studies have analyzed the limitations of static IO models in assessing economic losses, particularly regarding their architectural and data assumptions. Seminal literature in the field has argued that IO architecture is sensitive to aggregation levels and tends to overestimate losses by as much as 41% (Rose & Casler, 1996; Rose & Liao, 2005). Indeed, the lack of substitution effects, price elasticity, and production recapture leads static IO models with high outputs to overestimate more than those with more temporal granularity by nearly 70% (Rose & Wei, 2013). Moreover, several studies have noted IO modeling's sensitivity to input parameters and model architecture design because of the assumptions of production bottlenecks, inventory buffers, and lack of adaptive characteristics. Moreover, literature indicates that static IO models like the Leontief and Ghosh matrices are highly sensitive to modeling assumptions and model architecture, even doubling estimates when augmenting assumptions about inventory substitutability (Hallegatte, 2008, 2012).

In all, existing literature affirms that IO analysis, including Leontief and Ghosh, critical input, and inoperability IO methods, have been employed to study the impacts of systemic shocks with upstream and downstream economic effects on macroeconomic supply chains. These sources illustrate that IO architecture is an established approach for considering inter-industry linkages and analyzing the impacts of power outages in the context of natural hazards. Therefore, studying the sensitivity and output variations from modeling architecture selection and data parameterization is of valid concern for analysts to understand economic estimates of GDP losses for policy implications and risk analysis.

# Section 3: Methodology

In this section, the methodology for the paper is presented. First, we discuss three methods for how the initial final demand shock for the model is estimate. Next, we discuss the three methods to parametrize this shock in IO analysis to produce estimates regarding changes in final output.

## Section 3.1: Final Demand Shock Parameterization

### Section 3.1.1: The Household Interruptions Calculation

The household interruptions parameterization begins by collecting data from news sources, government reports, and other research journals about the scope and affected areas of power outages across the United States. Table 1 provides sources and data for all case studies analyzed within the context of this paper.

| Event | Region | Approx. Household Interruptions | Duration | Sources |
|---|---|---|---|---|
| Hurricane Ian Blackouts (2022) | Southeast | 2.7 Mn | 12 days | (USDE, 2022r) |
| 2021 Texas Blackouts | Texas | 13.5 Mn | 14 days | (Flores et al., 2023) |
| Tropical Storm Isaias Blackouts (2020) | Northeast | 6.4 Mn | 7 days | (DOE, 2020; *Major Power Outages Events*, 2025; O'Neill, 2020; Samenow et al., 2020; Zaveri & Tully, 2020) |

*Table 1: Outage case studies evaluated in the paper and sources for household data*

Next, interrupted consumer hours of power are calculated using real-world data from the sources in Table 1. In Figure 1, consumer outages per hour throughout each blackout are presented. For cases where such data is not obtainable, numbers are derived and extrapolated from news, government, and third-party reports, where applicable.

After the impacted consumer hours are calculated, we estimate the impacted value of final demand. The EORA model gives the US final consumer demand for electricity, gas and water as US $533 Bn (1.78% of US GDP). Thus, dividing by the US population and the number of hours in a year, the final consumer demand for utilities is US $0.182 per consumer hour, assuming homogeneous coverage. We note that informal estimates report that a kWh of electricity use, on average, costs around US $.19 per consumer hour, which contributes to the validation of this model-based estimate (Casey McDevitt, 2025).

It is appropriate to assume homogeneity because all blackouts examined in this paper span multiple days, so all available hours are uniformly likely to be interrupted. Depending on data availability, it may be necessary to adjust with respect to kilowatt-hours per consumer hour, which may increase in months with extreme temperatures. Multiplying the total number of consumer hours impacted by US $.182 yields the relevant change in annual final demand. Other analyses of the cost of power interruptions have incorporated similar parameterization techniques, and the number of customers without power has been identified as a data source for quantifying the cost of power outages (Ghodeswar et al., 2025).

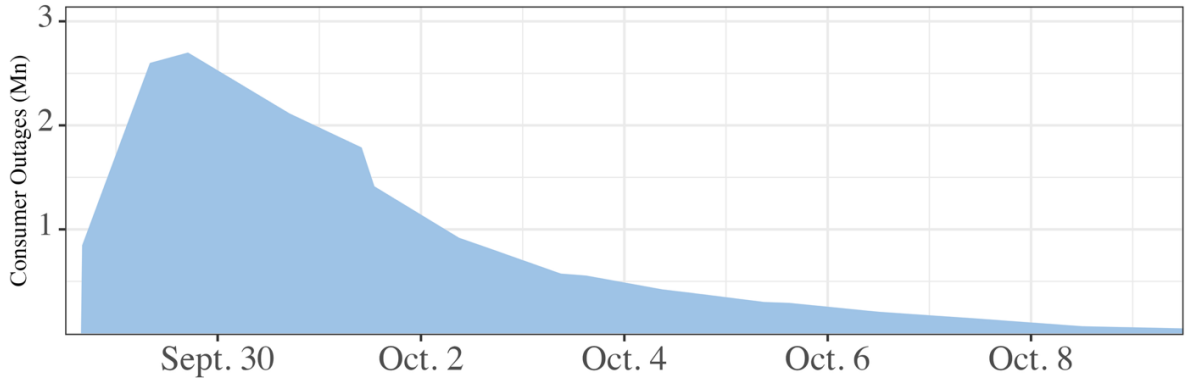

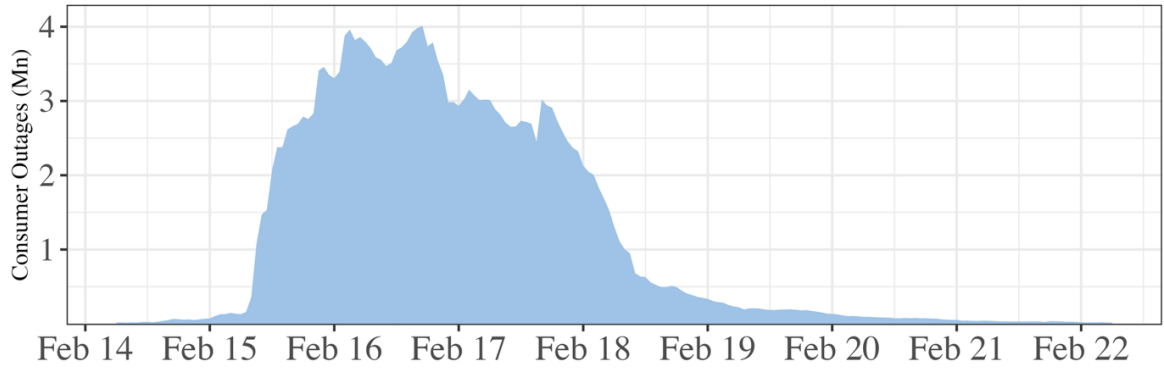

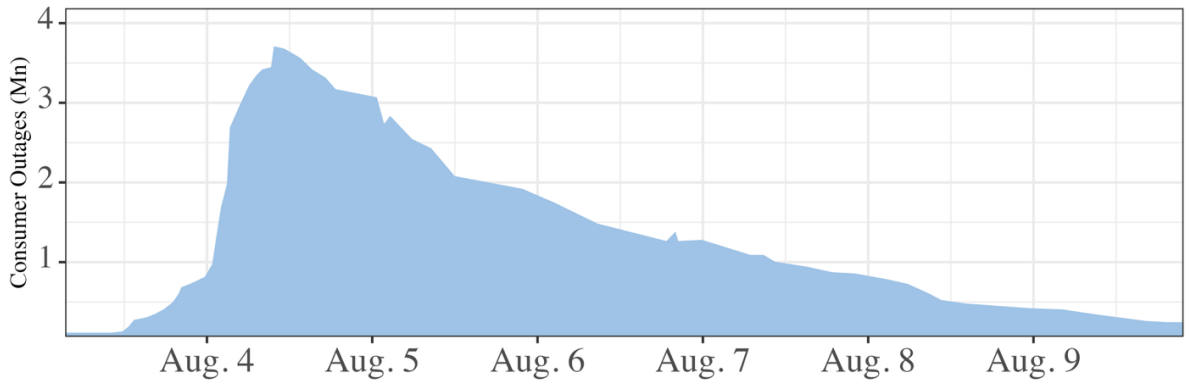

*Figure 1: [A] Consumer outages after Hurricane Ian (USDE, 2022), [B] Consumer outages in the 2021 Texas Blackout (Flores et al., 2023), [C] Consumer outages after Hurricane Isaias (DOE, 2020)*

*Section 3.1.2: The Kilowatts of Energy Lost Calculation*

As in the previous section, the final demand shock calculation relies on previous studies and news sources to estimate the kilowatt-hours (kWh) lost. Table 2 provides sources and data for all case studies analyzed within the context of this paper.

| Event | Region | Estimated kWh Lost | Duration | Sources |
|---|---|---|---|---|
| Hurricane Ian Blackouts (2022) | Southeast | $3.95 \cdot 10^8$ | 10 days | (USDE, 2022) |
| 2021 Texas Blackouts | Texas | $3.97 \cdot 10^{9}$ [1] | 14 days | (King et al., 2021) |
| Tropical Storm Isaias Blackouts (2020) | Northeast | $3.51 \cdot 10^8$ | 7 days | (DOE, 2020) |

*Table 2: Outage case studies evaluated in the paper and sources for household data*

Subsequently, the estimated kilowatt-hours interrupted by the natural disaster are divided by the net generation of electricity in the United States, which the US Energy Information Administration estimates to be 4.25 petawatt-hours (4,251.62 billion kWh) (EIA, 2024b). This figure represents the estimated percentage of energy generation lost due to the outage.

After this percentage is estimated, the model assumes that the final demand for electricity decreases commensurately, as the outage is precipitated by the inability to access electricity. As IO modeling permits parametrizing disruptions through final demand shocks to consumers, we normalize this kWh lost with respect to the residential end use of electricity (EIA, 2025). Finally, the final demand shock is calculated by augmenting the final demand commensurately with the percentage change in electricity output during the storm.

---

[1] Due to data-constraints, extrapolation was necessary to estimate the kWh lost during the blackout for the 4 days unreported in the data. Extrapolation was performed with respect to satellite luminosity data.

Similar methodologies focus on quantifying economic losses by analyzing electricity disturbances in the context of value of load lost analysis, such as Thomas & Fung (2022) and Hashemi et al. (2018).

*Section 3.1.3: Satellite Luminosity Calculation*

Some studies have employed satellite luminosity methods to estimate the final demand shock of electricity in affected regions. For this study, we employ the VIIIRS Nighttime Luminosity layers provided by NASA (Earth Science Data Systems, 2024). Specifically, we use the VNP46A2 dataset to access Gap-filled luminosity layers after Day/Night Correction (DNB-BRDF Corrected) to remove confounding. After accessing this data, we process it by masking, clipping, and normalizing the data. After processing, the data yields a map presenting luminosity layers, such as presented in Figure 2 [A]:

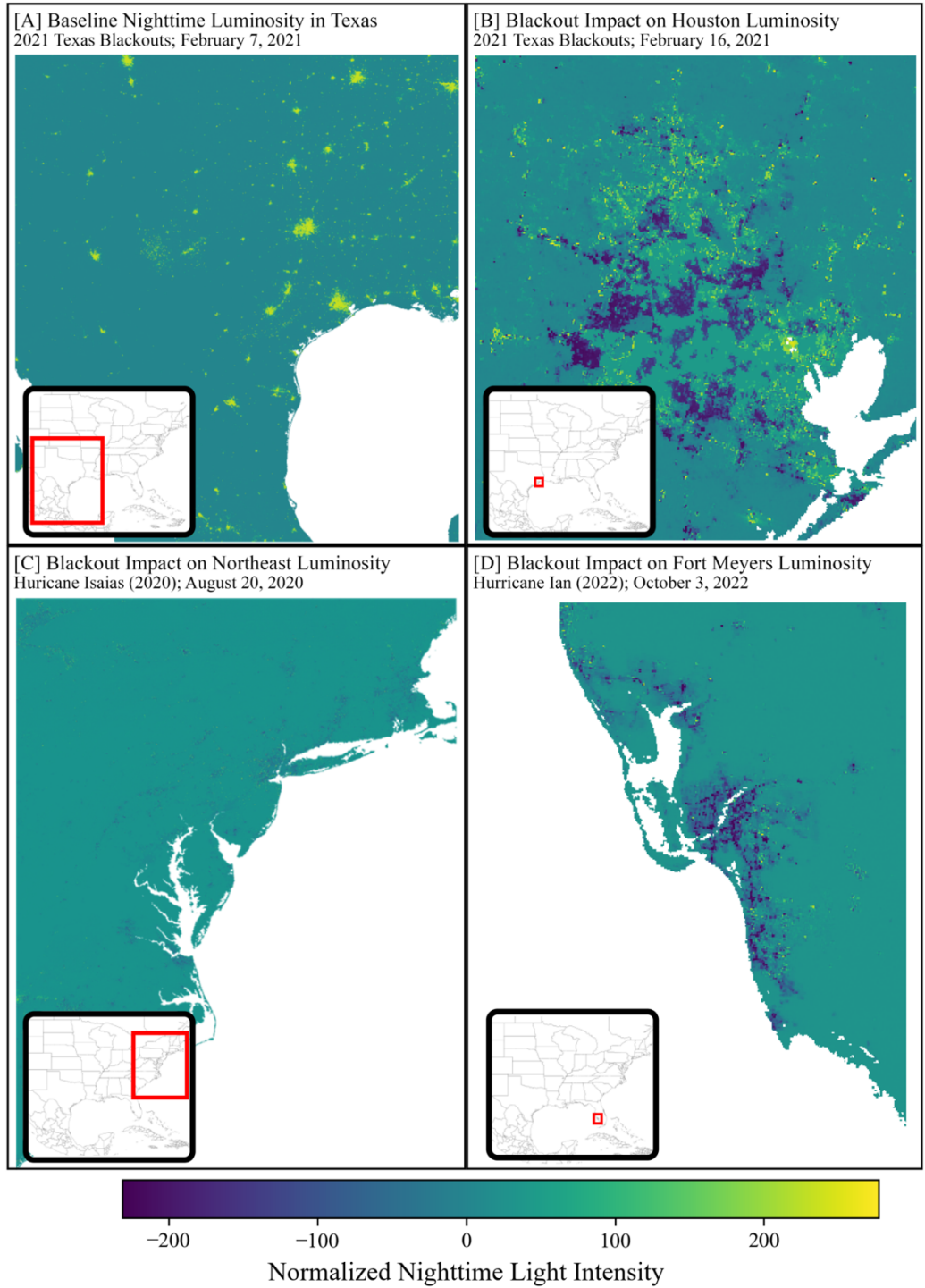

Figure 2: [A] Baseline luminosity layer for Texas, February 7, 2021, [B] Difference in absolute luminosity for February 7 and February 16, during the 2021 Texas Winter Storm, [C] Difference in nighttime luminosity between August 4, 2020 and July 31, 2020, [D] Normalized nighttime luminosity difference after Hurricane Ian on October 2, 2022

When subtracting by coordinate-level luminosity, we find the difference to be the measure of luminosity lost during the storm. We note this is a mild technical assumption, as the blackout was the only significant luminosity-diminishing event in this time frame, as the data has been corrected for snow and cloud cover. Consequently, our analysis yields plots like Figure 2 [B], which represent the difference in luminosity in the Houston metro area between February 7 and February 16.

To estimate the final demand impact, we first assume homogeneity of baseline electricity usage, a technical assumption necessary due to data constraints. Consequently, we compare lost luminosity over a region with the baseline luminosity to generate a percentage of lost luminosity. Then, we assume that the percentage change in electricity final demand is commensurate with the percentage change in luminosity, which is an appropriate assumption because electricity final demand comprises a significant contribution to luminosity through lighting homes and businesses. A final demand shock is subsequently calculated by multiplying this estimated percentage change in luminosity by the final electricity sector demand.

Nighttime light has been employed as a parameterization technique for a plethora of economic loss studies (Chor & Li, 2024; Del Castillo et al., 2025; Fratesi et al., 2025; Zhao et al., 2025). For example, one study that employed similar methodologies includes Mitsova et al. (2024), which used nighttime VIIRS data to parametrize the shock associated with the blackouts after Hurricane Michael in 2018. Moreover, nightlight luminosity data has been used to measure power system recovery after tropical cyclones and estimate economic impacts of Caribbean Hurricanes (Blackburn et al., 2021; Mo et al., 2025).

## Section 3.2: Input-Output Methodology

For the purposes of IO analysis, this paper will rely on the EORA model using data from 2017 (Lenzen et al., 2012, 2013). All augmentations described below will be applied to the Leontief and Technical Coefficients matrix of the EORA database unless otherwise specified.

### Section 3.2.1: The Leontief and Ghosh Method

To calculate the macroeconomic impact of lost consumer hours for households, we follow the IO procedure established by Miller & Blair (2009) for the Leontief and Ghosh modeling techniques.

This paper uses the Leontief model to analyze the upstream effects of outages. The EORA technical coefficients matrix $A$ and final demand vector $F$ are taken from the EORA database. Then, a Leontief IO matrix is calculated as follows, where $x$ represents the final output:

$$x = (I - A)^{-1}F$$

The Leontief IO matrix is denoted as $L = (I - A)^{-1}$. We can find the change in final demand by subtracting the impacted consumer-hours of demand from the electricity, gas and water sector in the final demand vector. Where $x'$ and $F'$ represent the new output and final demand after the shock:

$$(x - x') = (I - A)^{-1}(F - F')$$

$$\Delta x = (I - A)^{-1}\Delta F$$

$\Delta x$ represents the change in final output due to the outage's impact on industries upstream of the utilities sector. For downstream analysis, we use the Ghosh supply-side modeling procedure as recently analyzed by Altimiras-Martin (2024).

Using the diagonalized final output vector $\hat{x}$ and technical coefficients matrix $A$ given by the EORA database, we construct the direct allocations matrix $B$ as described in Miller & Blair (2009)

$$\widehat{x^{-1}}A\hat{x} = B$$

Then, using a similar procedure as the above, we calculate the Ghosh IO supply-side matrix $G$:

$$G = (I - B)^{-1}$$

We use this matrix to calculate the downstream changes in final output from changes in value-added as a consequence of the power outage:

$$x' = v'G$$

When combining the upstream and downstream changes in output, we must be cognizant of possible issues in double-counting. We resolve this issue by first separating the upstream indirect impact, the downstream indirect impact, and the direct impact from both calculated output changes, and adding the upstream indirect impact and the downstream indirect impact to find a total indirect impact. A similar methodology, combining Leontief and Ghosh matrices, was employed to study the impact of energy shortages in Pakistan (Rani et al., 2023).

*Section 3.2.2: The Critical Input Model*

Some research groups consider electricity to be a critical input good, implying that supply chains or manufacturing processes are dependent on electricity to function in the short run. Consequently, a power outage would temporarily halt output completely, as firms expect power restoration and do not shift to new input goods until the long run. The analysis assumes that all households are energy consumers, so an x% decrease in electricity final demand would create a commensurate x% interruption in final demand across the market. Similar assumptions were employed in modeling throughout the literature review, including Inoue et al. (2023).

When receiving the final demand shock discussed earlier in the paper, the model applies the same % shock to all sectors to model a short-run interruption of output capability. Such methodology is employed by studies such as Oughton et al. (2017), which explores the economic impact of blackouts as a result of extreme weather events. Similar assumptions have been mentioned in literature, as the relationship between electricity and electronic sectors cannot be fully explained by the technical coefficients in standard IO analysis because manufacturing output depends almost completely on electricity (Kelly, 2015; Thomas & Fung, 2023).

*Section 3.2.3: The Inoperability IO Matrix*

To estimate the macroeconomic impact of diminished production capability for electricity, we follow standard IO procedures established in literature for the inoperability IO modeling technique.

First, an interdependency matrix $A^*$ is calculated as follows, where $\hat{x}$ represents the diagonalized output of each constituent sector of the economy, as calculated by the EORA database, and $A$ represents the Leontief technical coefficient matrix:

$$A^* = \hat{x}^{-1} \cdot A \cdot \hat{x}$$

Next, a perturbation vector $c^*$ is calculated from final demand reductions relative to original output:

$$c^* = \hat{x}^{-1} \cdot (\Delta x)$$

Then, an inoperability vector $q$ is calculated to quantify disruptions across sectors of the macroeconomy, where $I$ represents an identity matrix with the same dimensions as the technical coefficients matrix $A$:

$$q = (I - A^*)^{-1} \cdot c^*$$

Finally, we use the inoperability vector $q$ to estimate upstream economic losses by calculating the loss of output associated with the diminished production:

$$x' = \hat{x} \cdot q$$

Similar inoperability-based methodology has been employed by studies like Bhattacharyya & Hastak (2022), which investigates the macroeconomic impact of the 2021 Texas Blackouts by focusing on sectoral inoperability.

# Section 4: Results

In this section, the results are reported for the research questions defined in the introduction. In Section 4.1, we report the findings of the model on several case studies of large-scale power outages in the US and study the GDP impacts. In Section 4.2, we compare the alternative data parameterization methods with empirical and observed losses.

## Section 4.1: What are the GDP impacts of several large-scale power outages in the United States?

### Section 4.1.1: Hurricane Ian (2022) Blackouts

On September 28, 2022, Hurricane Ian made landfall near Fort Myers, Florida as a Category 4 hurricane, creating extensive damage (Hauptman et al., 2024; C. Wang et al., 2024). By the next morning, over 2.6 million households had lost power as a result of catastrophic damage to energy infrastructure across Florida, the Carolinas, Georgia, and Virginia. Blackouts would persist for a week and a half as power was restored (LeComte, 2023; Lee et al., 2024). Data sourced from the US Department of Energy reports that there were over 237 million consumer hours of electricity interrupted by the storm. A complete breakdown is depicted in Figure 1 [A], as well as alongside day-by-day sources in Table 12 in the appendix. Satellite luminosity layers detected severe blackouts in the Fort Myers area in Southern Florida, depicted in Figure 2 [D].

Consequently, the household interruption calculation method estimates that the direct impact of lost electricity demand could be up to US $43.2 Mn. Estimating that the average Florida household uses 1,200 kWh per month in the summer, the kWh lost calculation estimates that the direct shock to electricity demand potentially approached US $99.1 Mn (Foley, 2024). The satellite luminosity parameterization technique estimates that the direct shock to the blackout could have been as high as US $161 Mn.

The domestic model results are reported below in Figure 3 and Table 3. The household interruptions, kWh lost, and satellite luminosity calculations yield a direct shock of US $43.2 Mn, US $99.1 Mn, and US $161 Mn, respectively.

|  | Leontief and Ghosh Method | Critical Input Method | Inoperability IO Method |
| --- | --- | --- | --- |
| Household Calculation | US $76.5 Mn | US $3.78 Bn | US $29.3 Mn |
| kWh Lost Calculation | US $176 Mn | US $8.68 Bn | US $67.3 Mn |
| Satellite Luminosity Calculation | US $286 Mn | US $14.1 Bn | US $109 Mn |

*Table 3: Domestic indirect losses of the three methods (columns) and three parameterization methods (rows) from the Hurricane Ian (2022) Blackouts*

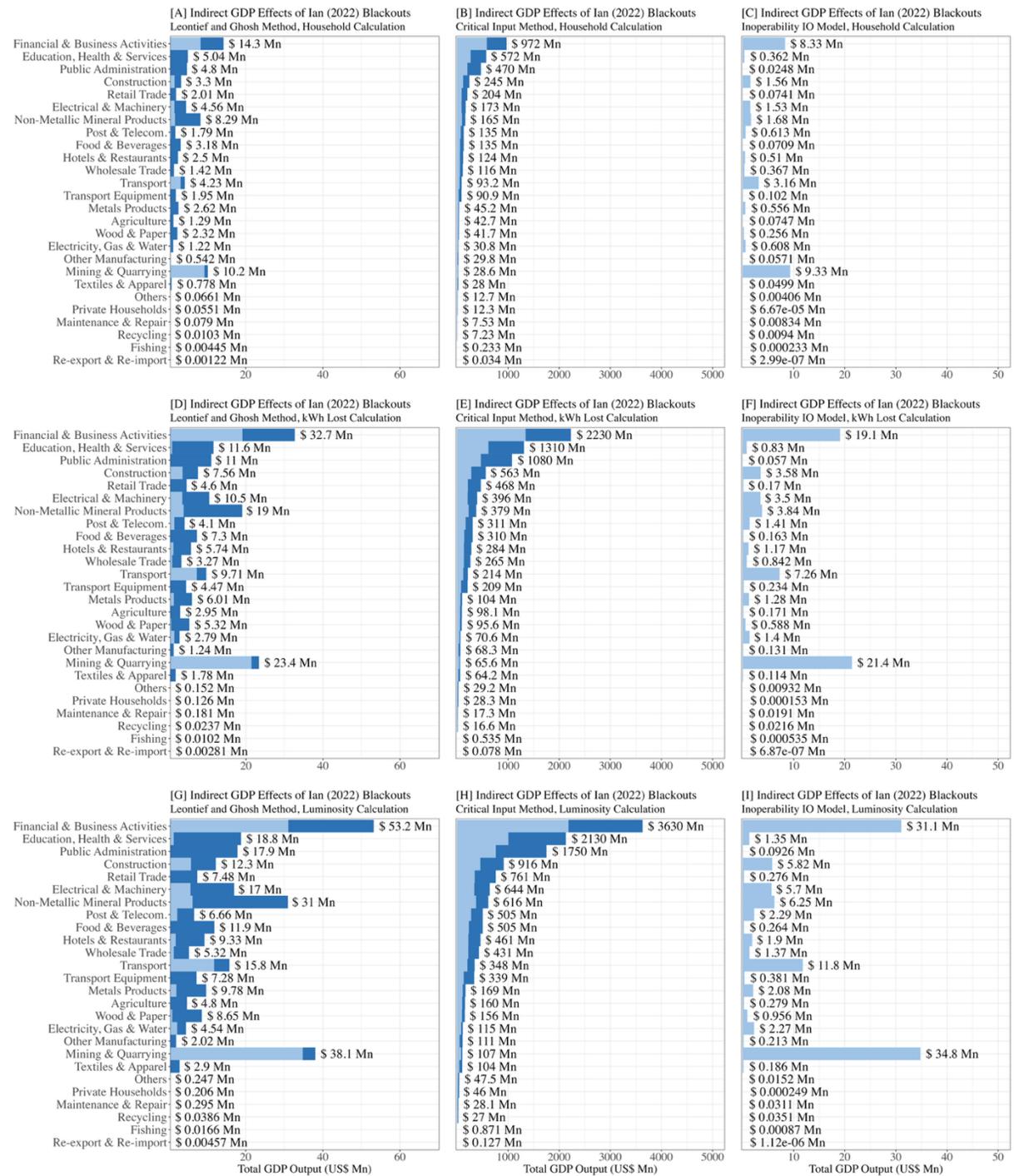

*Figure 3: Exploring the domestic indirect GDP loss estimates of the three methods (columns) and three parameterization methods (rows) from the Hurricane Ian (2022) Blackouts*

The global model results are reported below in Figure 4 and Table 4.

|  | Leontief and Ghosh Method | Critical Input Method | Inoperability IO Method |
|---|---|---|---|
| Household Calculation | US $88.5 Mn | US $4.09 Bn | US $35.9 Mn |
| kWh Lost Calculation | US $203.1 Mn | US $9.39 Bn | US $82.3 Mn |
| Satellite Luminosity Calculation | US $330 Mn | US $15.3 Bn | US $134 Mn |

*Table 4: Global indirect losses of the three methods (columns) and three parameterization methods (rows) from the Hurricane Ian (2022) Blackouts*

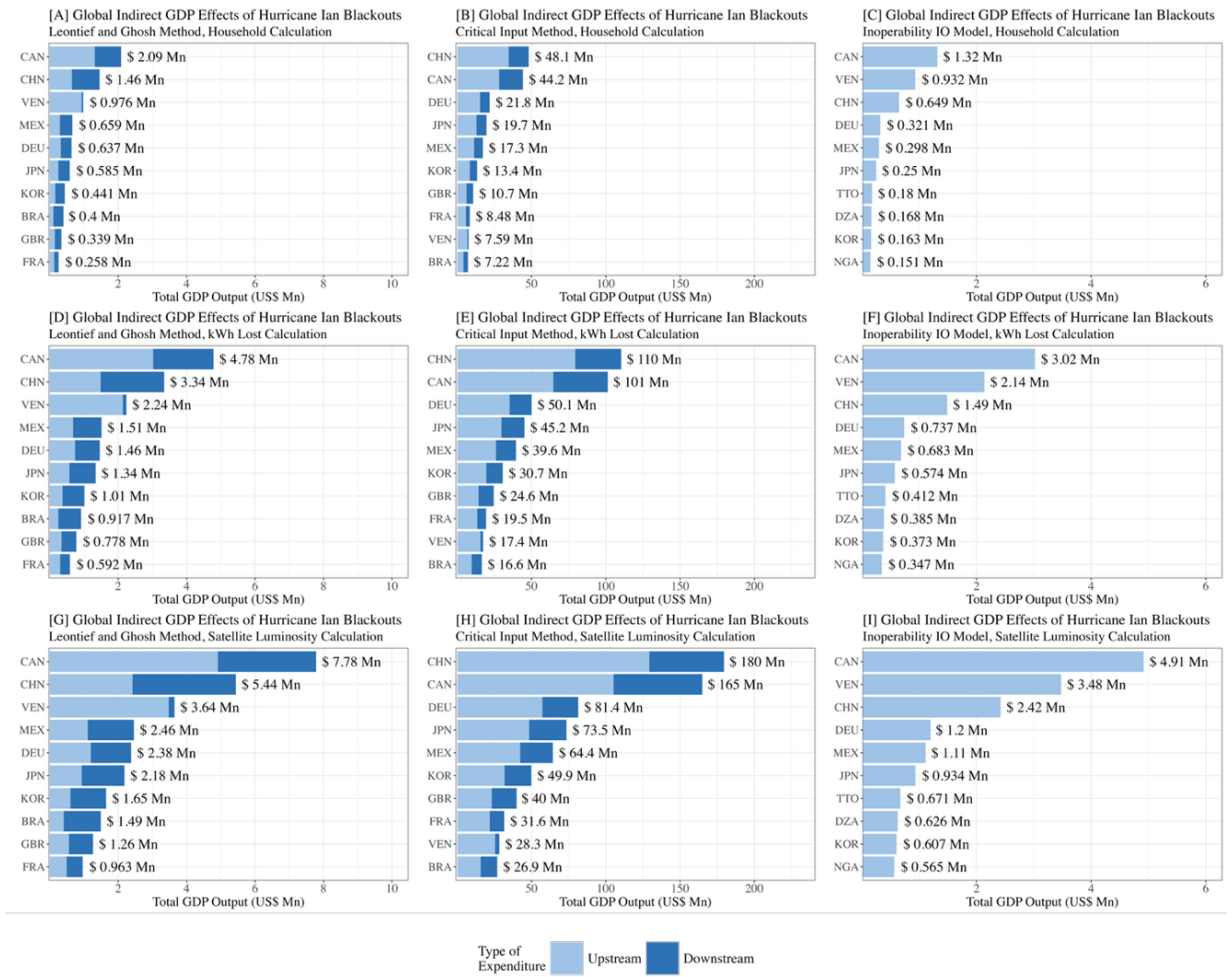

*Figure 4: Exploring the global indirect GDP loss estimates of the three methods (columns) and three parameterization methods (rows) from the Hurricane Ian (2022) Blackouts*

*Section 4.1.2: 2021 Texas Blackouts*

In February 2021, three severe weather storms stimulated by a La Niña pattern and a record-breaking severe winter season triggered a significant failure in Texas energy infrastructure and caused widespread blackouts in the region (Vera, 2021). The storm would leave approximately 4.5 million households without power for as long as two weeks (Busby et al., 2021). One study found that 227 million consumer hours were disrupted, illustrated by the time-series graph in Figure 1 [B] (Flores et al., 2023).

The household interruptions calculation yields a final demand shock of US $66.7 Mn. In months with extreme temperatures, Texans use an estimated 1,200 kWh in a month (Lumley, 2025). Consequently, the kWh-lost calculation method reports a final demand shock of US $192 million. Finally, the satellite luminosity calculation method reports a final demand shock of US $148 million. The domestic indirect losses are depicted in Figure 5 and Table 5.

|  | Leontief and Ghosh Method | Critical Input Method | Inoperability IO Method |
| --- | --- | --- | --- |
| Household Calculation | US $118 Mn | US $5.84 Bn | US $45.3 Mn |
| kWh Lost Calculation | US $340 Mn | US $16.8 Bn | US $130 Mn |
| Satellite Luminosity Calculation | US $262 Mn | US $13.0 Bn | US $100 Mn |

*Table 5: Domestic indirect losses of the three methods (columns) and three parameterization methods (rows) for the 2021 Texas Blackouts*

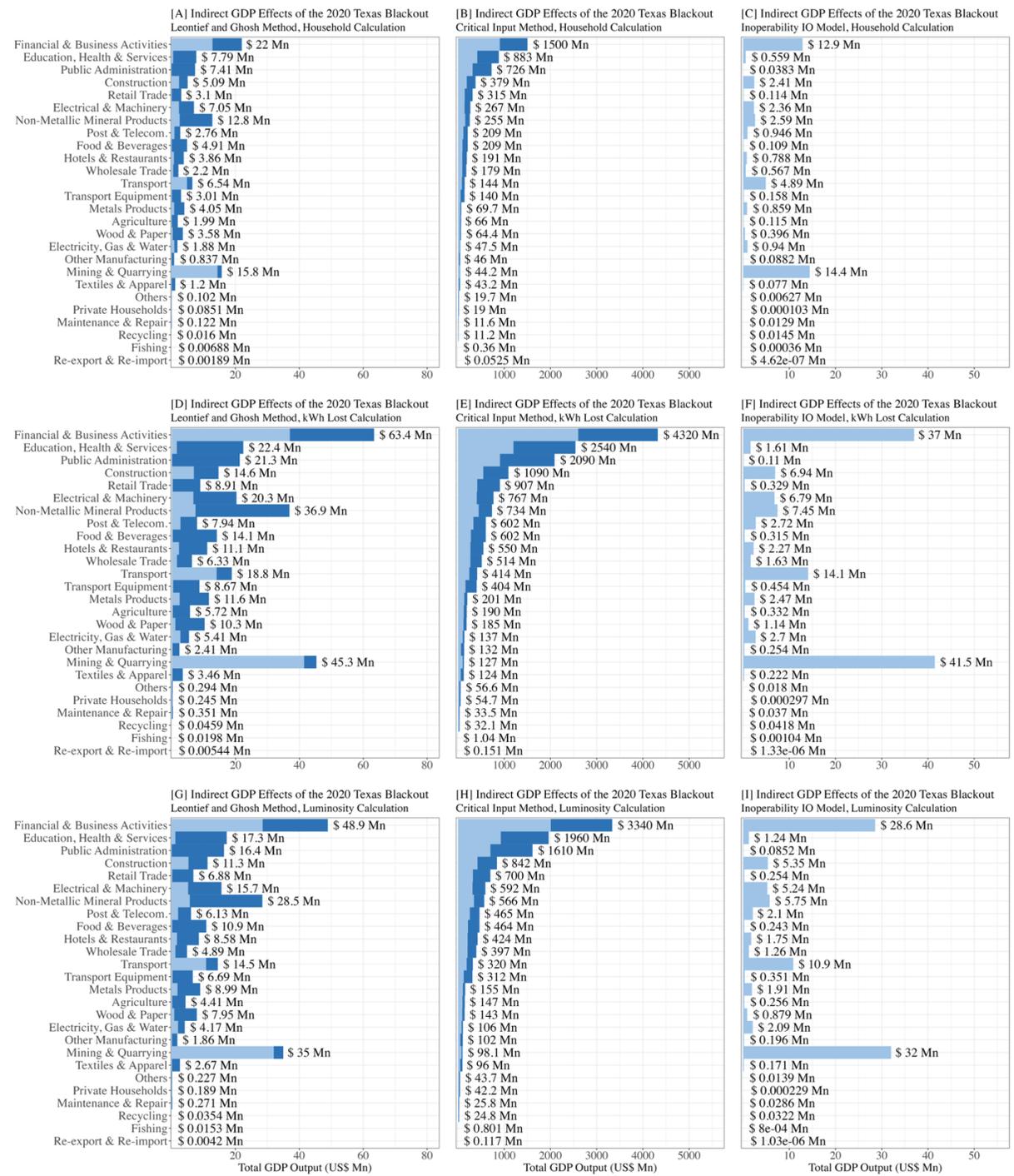

*Figure 5: Exploring the domestic indirect GDP loss estimates of the three methods (columns) and three parameterization methods (rows) for the 2021 Texas Blackouts*

The global model outputs are reported in Figure 6 and Table 6. The household interruptions, kWh lost, and satellite luminosity calculations yield a direct shock of US $66.7 Mn, US $192 Mn, and US $148 Mn.

|  | Leontief and Ghosh Method | Critical Input Method | Inoperability IO Method |
|---|---|---|---|
| Household Calculation | US $137 Mn | US $6.32 Bn | US $55.5 Mn |
| kWh Lost Calculation | US $394 Mn | US $18.2 Bn | US $159 Mn |
| Satellite Luminosity Calculation | US $304 Mn | US $14.0 Bn | US $123 Mn |

*Table 6: Domestic indirect losses of the three methods (columns) and three parameterization methods (rows) for the 2021 Texas Blackouts*

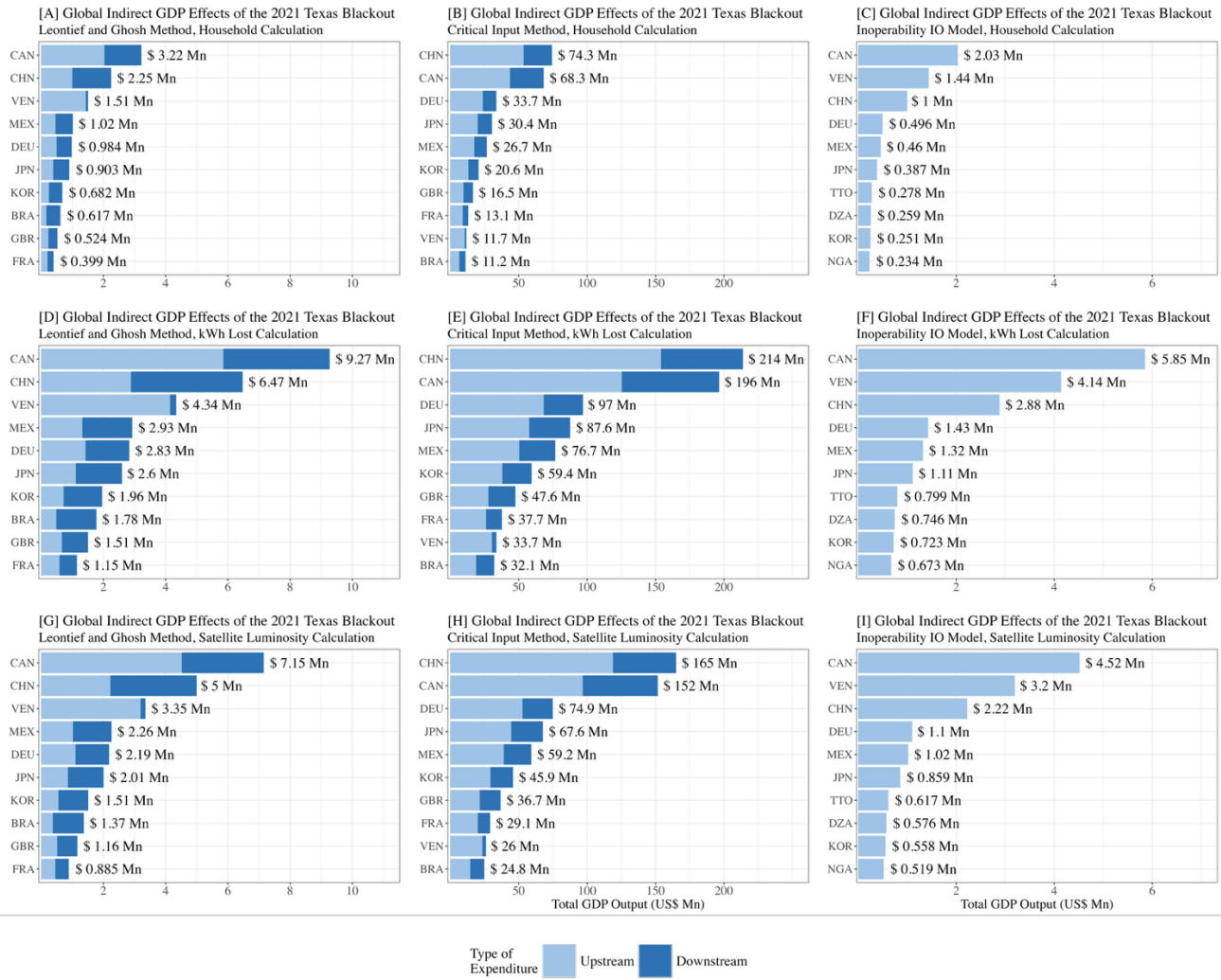

*Figure 6: Exploring the global indirect GDP loss estimates of the three methods (columns) and three parameterization methods (rows) for the 2021 Texas Blackouts*

*Section 4.1.3: Tropical Storm Isaias (2020) Blackouts*

As part of the record-breaking 2020 Atlantic hurricane season, Category 1 Hurricane Isaias instigated considerable blackouts after damaging power infrastructure in New Jersey, New York, Connecticut, and other northeastern states (Arora & Ceferino, 2024; Birchler et al., 2025; Rouhana et al., 2025). Throughout this paper, the storm is referred to as Tropical Storm Isaias because it had weakened into a tropical storm by the time it reached the East Coast.

News sources, cited in Section 3.1.1, provided evidence for extreme rolling blackouts from August 3 to August 11. Figure 1 [C] reports consumer outages per hour throughout the blackouts. Figure 2 [C] documents a blackout across the Northeast from Maine to Georgia, but most prominent in New Jersey, New York, and Connecticut. News reports indicate the peak outage included disruptions to more than 3.6 million outages on August 4. From other sources and a DOE situation report that provided data on consumer outages per hour, we can estimate that 281 million consumer-hours of power were lost. We note that, while peak outages were lower than the Texas 2024 Blackout, recovery efforts progressed more slowly, likely due to extensive damage from the hurricane-force winds. Consequently, this parameterization method reports a final demand shock to the electricity sector of US $51.1 Mn.

Assuming that a northeastern household uses 150 kWh per week (Agway Energy Services, 2024), the model estimates the kWh-lost calculation yields a loss of US $91.1 Mn. The satellite luminosity calculations found a final demand shock of $142.2 Mn.

Below, in Figure 7 and Table 7, the domestic model results are reported. The household interruptions, kWh lost, and satellite luminosity calculations yield a direct shock of US $51.1 Mn, US $91.1 Mn, and US $142.2 Mn.

|  | Leontief and Ghosh Method | Critical Input Method | Inoperability IO Method |
|---|---|---|---|
| Household Calculation | US $90.5 Mn | US $4.47 Bn | US $34.7 Mn |
| kWh Lost Calculation | US $161 Mn | US $7.98 Bn | US $61.9 Mn |
| Satellite Luminosity Calculation | US $252 Mn | US $12.4 Bn | US $96.5 Mn |

*Table 7: Domestic indirect losses of the three methods (columns) and three parameterization methods (rows) for the Tropical Storm Isaias (2020) Blackouts*

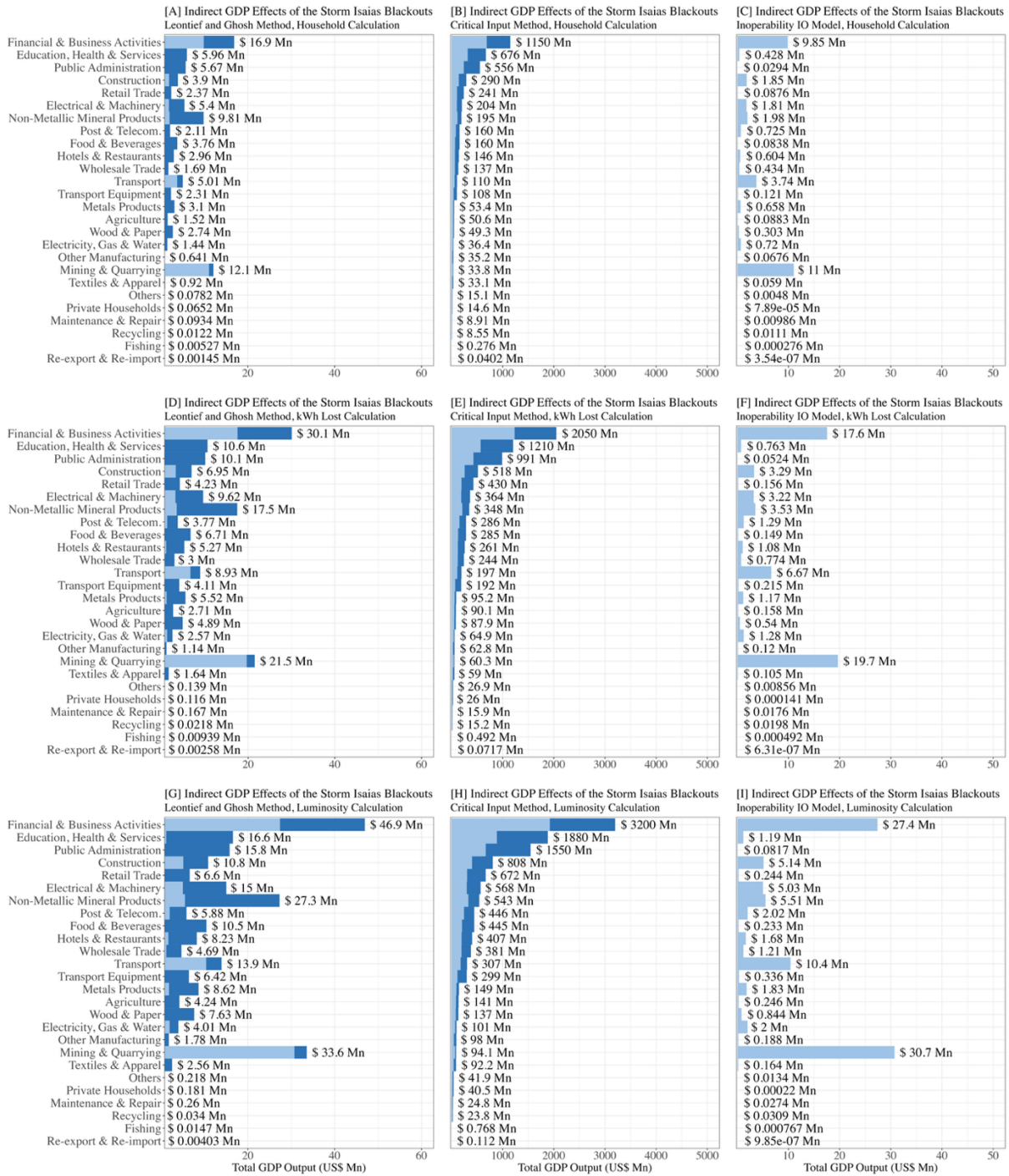

*Figure 7: Exploring the domestic indirect GDP loss estimates of the three methods (columns) and three parameterization methods (rows) for the Tropical Storm Isaias (2020) Blackouts*

Below, in Figure 8 and Table 8, the global model results are reported.

|  | Leontief and Ghosh Method | Critical Input Method | Inoperability IO Method |
|---|---|---|---|
| Household Calculation | US $104 Mn | US $4.82 Bn | US $42.5 Mn |
| kWh Lost Calculation | US $186 Mn | US $8.63 Bn | US $76.7 Mn |
| Satellite Luminosity Calculation | US $291 Mn | US $13.5 Bn | US $118 Mn |

*Table 8: Global indirect losses of the three methods (columns) and three parameterization methods (rows) for the Tropical Storm Isaias (2020) Blackouts*

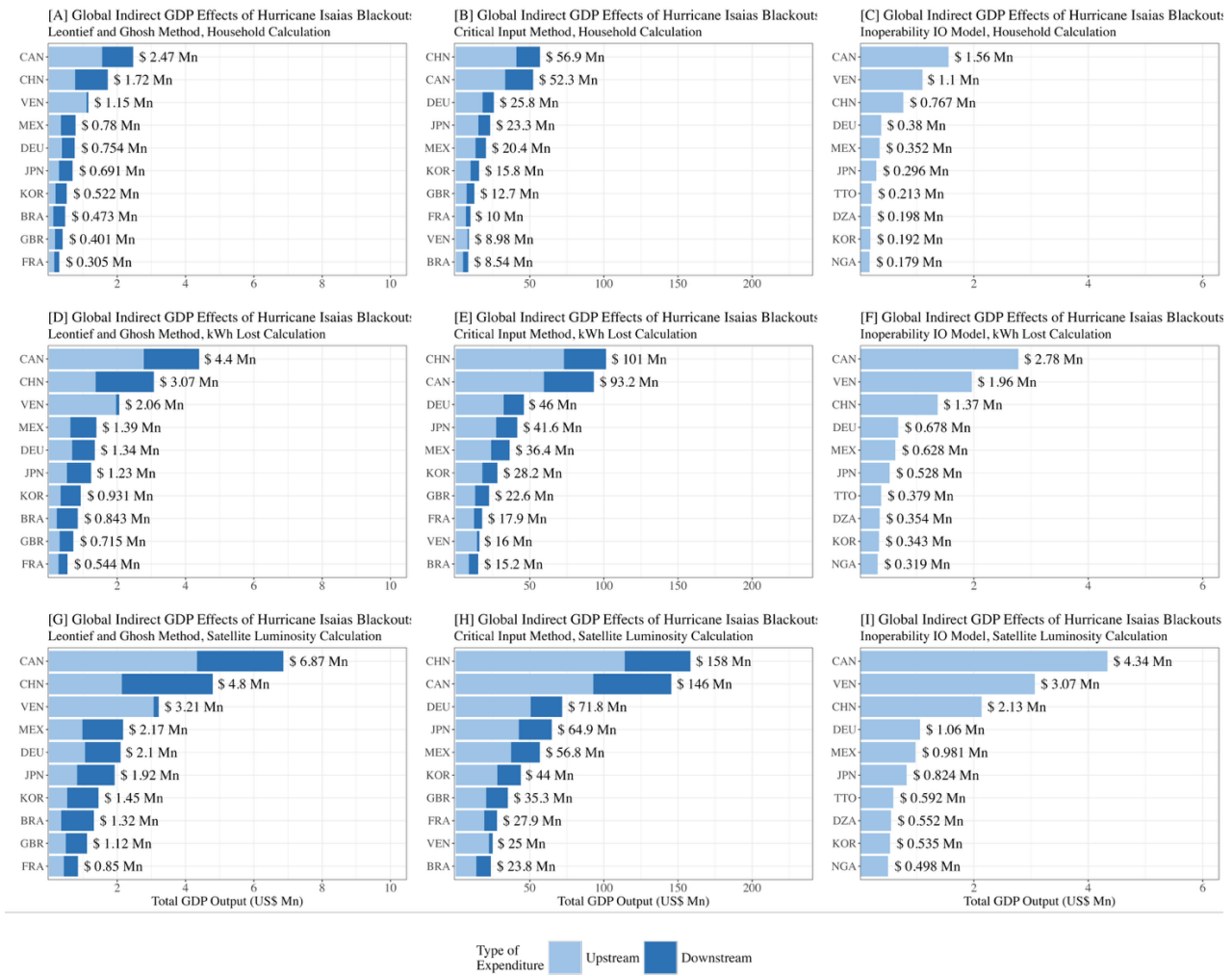

Figure 8: Exploring the global indirect GDP loss estimates of the three methods (columns) and three parameterization methods (rows) for the Tropical Storm Isaias (2020) Blackouts

Section 4.2: How do alternative data parameterization methods of natural hazard-induced outages compare with empirical and observed losses?

We investigate three case studies and find that the Hurricane Ian (2022) Blackouts, the 2021 Texas Blackouts, and Tropical Storm Isaias (2020) Blackouts could have created a total domestic loss of up to US $14.1 Bn, US $16.8 Bn, and US $12.4 Bn, respectively. Globally, the impact was projected to reach up to US $15.3 Bn, US $18.2 Bn, and US $13.5 Bn, respectively.

Summary statistics for the domestic and global loss estimates for the case studies are presented in Tables 9 and 10, respectively. The standard deviations are calculated across models and parameterization techniques.

|  | Hurricane Ian (2022) Blackouts (US $ Bn) | 2021 Texas Blackouts (US $ Bn) | Tropical Storm Isaias (2020) Blackouts (US $ Bn) |
|---|---|---|---|
| Mean total loss | 3.13 | 4.18 | 2.93 |
| Standard Deviation | 5.07 | 6.48 | 4.55 |
| Range | .0725 – 14.1 | .112 – 16.8 | .0432 – 13.5 |

Table 9: Domestic summary statistics for the Hurricane Ian (2022) Blackouts, the 2021 Texas Blackouts, and the Tropical Storm Isaias (2020) Blackouts.

|  | *Hurricane Ian (2022) Blackouts (US $ Bn)* | *2021 Texas Blackouts (US $ Bn)* | *Tropical Storm Isaias (2020) Blackouts (US $ Bn)* |
|---|---|---|---|
| Mean total loss | 3.29 | 4.41 | 3.09 |
| Standard Deviation | 5.49 | 7.00 | 4.93 |
| Range | .0791 – 15.3 | .122 – 18.2 | .0936 – 13.5 |

*Table 10: Global summary statistics for the Hurricane Ian (2022) Blackouts, the 2021 Texas Blackouts, and the Tropical Storm Isaias (2020) Blackouts.*

This analysis implies that choices in modeling architecture and data parameterization have a considerable influence on the magnitude of loss estimates. Differentiation across modeling architectures is understandable as each is designed to capture different measures of economic loss. However, differentiation across parameterization methods can only be explained by *a priori* economic assumptions, as, within a model, two parameterization methods measure the same outcome. This study finds that parameterization methods are highly variable and sensitive to the reliability, availability, and accessibility of data. This variability is evident when quantifying the range across data parameterization methods given modeling architectures.

With the modeling architecture held constant, the standard deviation of output across parameterization methods can be found in Table 11. Each standard deviation is displayed alongside the percentage of the mean of the model output across parameterization techniques.

|  | *Hurricane Ian (2022) Blackouts (US $ Bn, % of mean)* | *2021 Texas Blackouts (US $ Bn, % of mean)* | *Tropical Storm Isaias (2020) Blackouts (US $ Bn, % of mean)* |
|---|---|---|---|
| Leontief and Ghosh | .120 (38.9%) | .148 (33.4%) | .0936 (32.4%) |
| Critical Input | 5.59 (57.7%) | 6.02 (46%) | 4.33 (47.7%) |
| Inoperability IO | .049 (26.5%) | .0528 (21.2%) | .0379 (21.7%) |

*Table 11: Standard deviations of output across parameterization methods across modeling architectures for the Hurricane Ian (2022) Blackouts, the 2021 Texas Blackouts, and the Tropical Storm Isaias (2020) Blackouts.*

So, across the three storms, the data parametrization methods contributed to differences, on average, of 34.9% for the Leontief and Ghosh model, 50.5% for the critical input model, and 23.1% for the inoperability IO model. Consequently, we note that the output is sensitive to data parameterization methods.

We evaluate the performance of the nine combinations of data parameterization techniques and models by comparing results against previous studies of the macroeconomic impact of blackouts after natural disasters. One study found that the expected loss for the Texas 2021 Blackout was US $664 Mn (Bhattacharyya & Hastak, 2022). The National Oceanic and Atmospheric Administration estimates the blackouts cost up to US $27.2 Bn (Smith, 2025). Due to validation data limitations, we analyze the output for Tropical Storm Isaias and Hurricane Ian in the context of similar storms. A government report estimates that the outage costs due to Hurricane Ike (2008) could have been as high as US $24 Bn to $45 Bn, while the outage costs due to Hurricane Sandy (2012) could have been as high as US $14 Bn to $26 Bn (Executive Office of the President, 2013). Another study reported that Hurricane Sandy had a total impact of up to US $7 Bn (Alemazkoor et al., 2020).

Additionally, the storms created blackouts that coincidentally had similar consumer-hours affected, so we employ the Bhattacharyya & Hastak (2022) estimate of US $664 Mn indirect cost. We note that we cannot use estimates of the total cost of the storm for validation, as they are inflated, as they include infrastructure, housing, and property damage costs, which greatly outweigh the indirect economic cost of the blackouts.

As depicted in Figure 9, alternative parameterization techniques have an impact on the estimate of GDP loss as well as the choices in model architecture.

## [A] 2022 Hurricane Ian Blackout Validation Table

Leontief and Ghosh Model (LGM), Critical Input Model (CIM), and Inoperability IO Model (IIM)

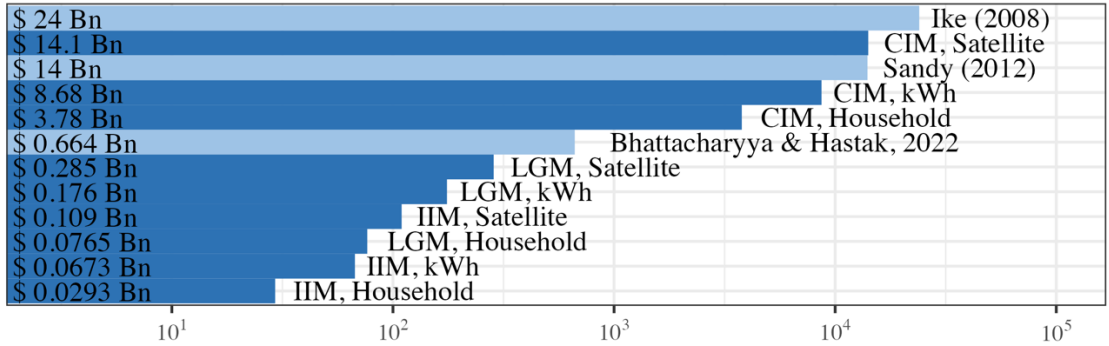

## [B] 2021 Texas Blackout Validation Table

Leontief and Ghosh Model (LGM), Critical Input Model (CIM), and Inoperability IO Model (IIM)

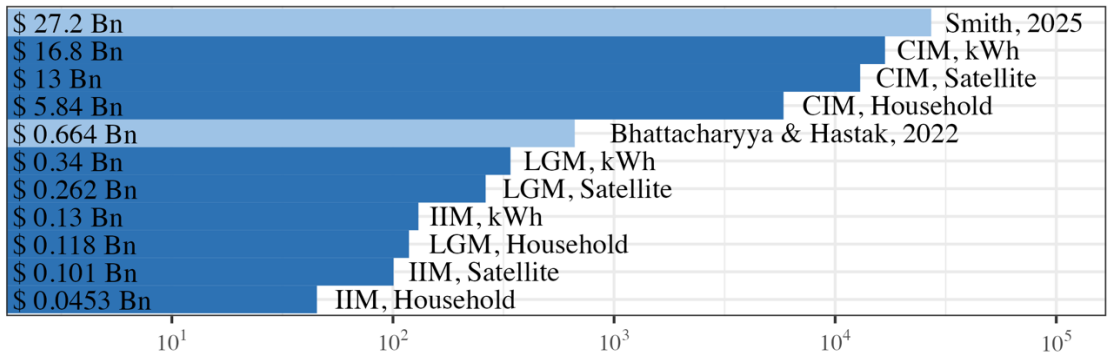

## [C] 2020 Hurricane Isaias Blackout Validation Table

Leontief and Ghosh Model (LGM), Critical Input Model (CIM), and Inoperability IO Model (IIM)

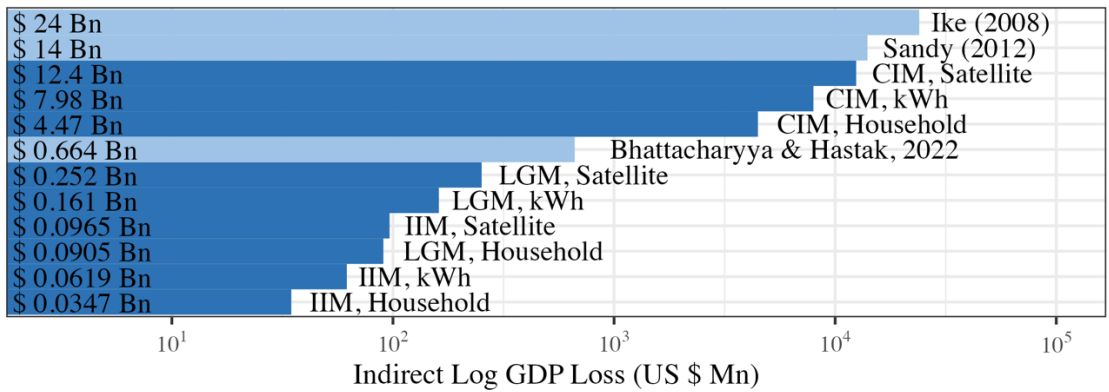

*Figure 9: Comparison data for domestic losses*

# Section 5: Discussion

In this section, a discussion will be undertaken to evaluate the results of Section 4 in the context of the research questions. We will also discuss the validity and the implications of the results.

## Section 5.1: What are the GDP impacts of several large-scale power outages in the United States?

We undertake three case studies and find that for Hurricane Ian (2022), the 2021 Texas Blackouts, and Tropical Storm Isaias (2020) could have created total global losses of up to US $15.3 Bn, US $18.2 Bn, and US $13.5 Bn, respectively. Domestically, the impact could have been as high as US $14.1 Bn, US $16.8 Bn, and US $12.4 Bn, respectively. Across all models and data parameterization methods, the disaster-induced blackouts had a mean GDP impact of US $3.13 Bn (standard deviation of US $5.07 Bn), US $4.18 Bn (US $6.48 Bn), and US $2.93 Bn (US $4.55 Bn).

These methods are designed to capture distinct dimensions of the macroeconomy, contributing to the wide range of economic loss predictions. The inoperability IO model only captures the upstream impact of the power outage, modeling a relatively small productivity shock on industries that rely on electricity as an input. In contrast, the Leontief and Ghosh method estimates both upstream and downstream impacts. The critical input method, which relies on the Leontief and Ghosh method, makes the additional assumption that a lack of electricity as an input good yields a commensurate lack of production. The critical input method is best suited for modeling short-term outages, as it does not incorporate restoration or behavioral shifts, such as the use of backup generators or re-routing methods. Therefore, the choice of modeling architectures will depend on the use case of future studies and the availability of suitable data.

Additionally, the critical input model is more sensitive to data parameterization, as it had a standard deviation of almost US $5.59 Bn for the Hurricane Ian (2022) blackout, compared to the Leontief and Ghosh model and the inoperability IO model, with US $120 Mn and US $49 Mn, respectively. The greater difference is generated by the critical input model, which maintains linear intersectoral relationships between inputs and GDP loss, while generating higher values. Therefore, IO models that generate quantitatively higher losses are considerably more sensitive to changes in data parameterization. Thus, the choice of modeling architecture introduces variability to GDP loss estimates, which is appropriate, as a choice of modeling architecture changes the scope of the model measurement.

The modeling methodologies also differ with respect to intersectoral linkages throughout the macroeconomy. The critical input model reports that the most-affected private sectors are Financial & Business Activities (US $4.3 Bn loss in the 2021 Texas Blackouts, kWh parameterization method; 0.109% of sectoral output), Education, Health & Services (US $2.54 Bn; .097%), and Construction (US $1.09 Bn; .081%), which is consistent with Anderson et al. (2007), which found that business services and financial institutions were the most impacted by a natural-hazard induced blackout. Whereas, the Leontief and Ghosh model found that the most-affected private sectors were Financial & Business Activities (US $64.4 Mn loss in the 2021 Texas Blackouts; kWh parameterization method; 0.00163% of sectoral output), Mining & Quarrying (US $45.3 Mn; 0.0116%) and Non-Metallic Mineral Products (US $36.9 Mn; .162%), which is consistent with Bhattacharyya & Hastak (2022), which found that the technical services and mining-related sectors were most impacted.

Given the limitations of Input-Output methodology, the numerical value output of the models is less reliable than insights into intersectoral linkages and other macroeconomic interactions.

Additionally, the scale of the impact, information regarding the flow of damages to industries, and magnitude of damage within industries (ordinally, as a percentage or ranking) should be taken away from the model, as the numerical value itself should be regarded as an upper bound to shock when abstracting from recovery or behavioral effects.

## Section 5.2: How do alternative data parameterization methods of natural hazard-induced outages compare with empirical and observed losses?

This study finds that the final output of Leontief and Ghosh, critical input, and inoperability IO models in the context of disaster-induced blackout macroeconomic modeling is highly sensitive to data parameterization methods and assumptions about the initial direct impact of blackouts. With the modeling architecture held constant, the standard deviation across parameterization models was 34.9% of the mean for the Leontief and Ghosh model, 50.5% for the critical input model, and 23.1% for the inoperability IO model. Indeed, this is consistent with previous literature, which noted that adjusting assumptions within static IO architecture can, at times, double the GDP results of the model (Hallegatte, 2008, 2012). Furthermore, our findings are consistent with existing literature that states that IO modeling's sensitivity to input parameters can, at times, create differences of up to 41% to 70% with respect to those with more temporal granularity (Rose & and Wei, 2013; Rose & Liao, 2005).

When comparing model output to real-world estimations and data, the selection of a suitable methodology can successfully capture different components of the consequences of a power outage. For the 2021 Texas Blackout, the critical input methodology matched estimates in the literature that studied the total indirect impact of the power outage. This could be because the critical input method accounts for the production disruptions that occur during large-scale natural disasters, which often cause significant infrastructure damage and necessitate restoration efforts to restart production. However, variations in parameterization techniques introduced additional variability. For the 2022 Isaias Blackout, the satellite parameterization method yielded an 11% difference according to validation results, while the kWh lost metric showed a variation of 43%. Indeed, because the critical IO method produces comparably high outputs, assumes constant returns to scale, and endogenizes homogeneous failure across production, it is highly sensitive to changes in the parameterization method and more prone to input-sensitive results.

Accordingly, this paper concludes that, within the context of the case studies involved, the variability introduced by data parameterization is an additional consideration in providing consistent estimators for the GDP impact of power outages. Indeed, this does not limit the insights and suggestions that the models provide about supply-chain resiliency or intersectoral impacts (as a percentage) but advises analysts not to give undue respect to the GDP loss, as it is highly sensitive to assumptions and data availability, reliability, and accessibility.

Notably, in some cases, parameterization techniques provided inputs that were twice as high. Thus, data limitations and research interests are valuable in choosing parameterization techniques. All three options have inherent limitations that should be considered by the research team when selecting a methodology. The household interruptions parameterization technique often requires an assumption of heterogeneity and is limited by a lack of reliable, accessible, or available data; the kWh lost parameterization methodology also has these limitations as well as requiring additional assumptions of linearity of power infrastructure failures to direct shock; and the satellite luminosity technique also assumes homogeneity and may capture other events in the time-window, although it should be noted that global data is readily available and reliable.

When analyzing parameterization techniques for model estimation, data accessibility, availability, and reliability are considerable limitations to modeling procedures. Due to VIIRS data being publicly accessible from NASA, the satellite luminosity parameterization does not face similar data limitations as the household interruption or kWh lost parameterization methods, which rely on news sources, survey data, or government reports. These methods may rely on approximation procedures, such as homogeneity across time and region, MRIO Input-Output data, and kWh usage, and may introduce additional variability into the procedure. Consequently, researchers studying the macroeconomic impact of natural hazard-induced outages are encouraged to use publicly available data for replication studies and validation, as this paper has noted that the GDP loss estimates are considerably sensitive (contributing to differences as high as 30% for the same storm and model) to choices of model architecture and data parameterization techniques. This variability can be reduced by relying on publicly available data rather than focusing on news, private, or government estimates that might be inaccessible or unreliable.

# Section 6: Conclusion

In this paper, we assessed the potential macroeconomic impacts of several widespread blackouts in the United States. We primarily evaluated three parameterization techniques to estimate the initial shock a power outage might have on the economy: household interruptions, kWh lost due to lack of generation, and satellite imagery luminosity data. Then, we evaluated three static IO models to produce several estimates of possible indirect upstream and downstream macroeconomic shocks: the Leontief and Ghosh model, the critical input model, and the inoperability IO model.

This paper evaluated three case studies: The Hurricane Ian (2022) Blackouts, the 2021 Texas Blackouts, and the Tropical Storm Isaias (2020) Blackouts. Although the model estimates vary with respect to modeling architecture, parameterization, and output interpretation, this paper estimates that the blackouts, as a consequence of the disasters, could have created a total global loss of up to US $15.3 Bn (with a mean of US $3.13 Bn and standard deviation of US $5.07 Bn), US $18.2 Bn (US $3.13 Bn; US $5.07 Bn), and US $13.5 Bn ($2.93 Bn; $4.55 Bn), respectively. Domestically, the impact could have been as high as US $14.1 Bn, US $16.8 Bn, and US $12.4 Bn, respectively.

We conclude that the decisions made by economic analysts about model architecture, data parameterization, and data reliability, accessibility, and availability introduce considerable sensitivity and create a wide range of estimates, creating variations of up to 30% within model architectures. The three models capture different elements of the macroeconomic shock, so variation across model architectures is expected and consistent with existing literature. However, this paper concludes that different data parameterization techniques introduced standard deviations of an average of 34.9% of the mean for the Leontief and Ghosh model, 50.5% for the Critical IO model, and 23.1% for the inoperability IO model. Consistent with existing literature, adjusting assumptions can, at times, double the GDP results of the model, especially within models with relatively higher losses.

As with any methodology or modeling process, there are relevant limitations. Primarily, the EORA MRIO model is static and does not account for the dynamic elements of the macroeconomy, such as resilience, recovery effects, and price fluctuations. Given that IO modeling assumes constant returns to scale and a stable macroeconomic supply chain, we cannot simulate supply and demand forces playing out in a market environment. Importantly, as discussed in the literature review, the output of an IO estimate is best treated as an upper bound. Consequently, losses may have been lower than those reported in the model. Double counting may have also affected these estimates because IO methods may count the increase of one sector without recognizing the subsequent loss of another from changing inputs. As IO modeling holds these losses exogenous, they are not reflected in the result. Similarly, the model does not incorporate balancing household budgets in response to external shocks due to the assumption of macroeconomic stability. To balance out the effect of double-counting, value-added and demand multiplier effects were considered independently and did not overlap in the presented results.

Future research could incorporate dynamic systems into this analysis, exploring how parameterization for dynamic systems impacts model performance and insights. Future studies could also compare dynamic models against static ones to evaluate performance. Additionally, investigating alternative disasters, such as floods, earthquakes, and tornados, would be worthwhile. Future research could also incorporate additional parameterization techniques and static model architectures in various case studies to further investigate the sensitivity of different models.

# Section 7: Appendix

| Date | Consumer Outages | Source |
|---|---|---|
| September 28, 4 PM | 846,000 | (USDE, 2022a) |
| September 29, 8 AM | 2,600,000 | (USDE, 2022b) |
| September 29, 5 PM | 2,700,000 | (USDE, 2022c) |
| September 30, 5 PM | 2,114,000 | (USDE, 2022d) |
| October 1, 10 AM | 1,787,000 | (USDE, 2022e) |
| October 1, 1 PM | 1,414,000 | (USDE, 2022f) |
| October 2, 9 AM | 919,000 | (USDE, 2022g) |
| October 2, 4 PM | 820,000 | (USDE, 2022h) |
| October 3, 9 AM | 575,000 | (USDE, 2022i) |
| October 3, 3 PM | 556,000 | (USDE, 2022j) |
| October 4, 9 AM | 423,000 | (USDE, 2022k) |
| October 4, 3 PM | 394,000 | (USDE, 2022l) |
| October 5, 9 AM | 302,000 | (USDE, 2022m) |
| October 5, 3 PM | 293,000 | (USDE, 2022n) |
| October 6 | 208,000 | (USDE, 2022o) |
| October 7 | 141,000 | (USDE, 2022p) |
| October 8 | 69,000 | (USDE, 2022q) |
| October 9 | 48,000 | (USDE, 2022r) |

*Table 12: Sources for daily consumer outages for Hurricane Ian*

# Section 8: Bibliography